\documentstyle[12pt]{article}
\begin{document}
\begin{titlepage}
\title{Renormalization Group Study of Sandpile on the Triangular Lattice.}

\author{Vl.V.~Papoyan\thanks{E-mail: vpap@theor.jinrc.dubna.su} \\[2mm]
and \\[2mm]
A.M.~Povolotsky\thanks{E-mail: povam@thsun1.jinr.dubna.su} \\[2mm]
{\small \sl Bogoliubov Laboratory of Theoretical Physics,} \\
{\small \sl JINR, 141980 Dubna, Russia.}
}
\date{}
\maketitle
\begin{abstract}

We apply the renormalization group approach  to the
sandpile on the triangular lattice. The only attractive fixed
point is found. The obtained fixed point height probabilities
are compared with numerical simulations.
The value of critical exponent of avalanche size distribution
is found to be $\tau=1.36$. The probabilities of the sand
transition are compared with the branching probabilities
of the spanning trees on the triangular lattice which are
also evaluated.
\end{abstract}
\thispagestyle{empty}
\end{titlepage}

\section{Introduction.}
The main idea of the renormalization group (RG)
approach to critical fenomena is
the calculation of parameters of a critical
state by a successive transition from
small-scale interactions to the large-scale ones.
It is possible due to
the vanishing of characteristic temporal and
spatial scales at the critical
point leading to temporal and spatial self-similarity. A
wide range of the systems exhibiting these properties
is unified by the
concept of Self-Organized Criticality (SOC) introduced
by Bak {\it et al}
\cite{BTW}. These systems are characterized by the
evolution into the critical
state free from fine tuning of any parameters.
The sandpile model was proposed
in the same work to illustrate the idea of the SOC.
This cellular automaton
was found to be rather a simple object of
investigations, although rich in
content. The sandpile model involves the set of
sites on which the function
of height is defined. Time after time being added
to a random site, "grains
of sand" cause the growth of its height. If the
height reaches some
critical value, the site topples reducing its height
and increasing the
heights of several neighboring sites. In turn, if the
heights of the
neighboring sites reach the critical value, they also
topple and so on. The
"sand" can leave the system on the boundary. After
evolving for some time
the system comes into the recurrent subset of all
its configurations
which is the
critical state. Further the successive topplings
between the stable states
called avalanche return the system into the critical
state each time
it leaves it.

The most successful RG approach to the description
of the sandpile on the
square lattice has been proposed by Pietronero {\it et al}
\cite{PVZ}. It is based
on the real space RG scheme driven by steady state
condition feedback. The
real space renormalization is performed according to the
standard block-transformation scheme.
The
feedback mechanism allows one
to take into account the strong correlations
appearing at the critical
point. Practically, this approach identifies
the dynamic and static
parameters of the model. While the static
configurations are described in
terms of stable and critical cell distribution,
the dynamic renormalized
variables characterize the toppling rule for
the critical cells. The
conservation law valid in the critical state makes it
possible to obtain the
feedback relation linking critical values of the
dynamic parameters and the
static configuration weights. The obtained
renormalization equation has a
unique attractive nontrivial fixed point. By
considering the avalanche as
a relaxation process at the definite scale
the exponent $\tau $ of the
distribution of the avalanche size was expressed
through the fixed point
parameters.

The next step in this direction was made by
Ivashkevich \cite{EVI}. While
previous approach dealt only with the critical
cell distribution,
the cells with different heights were introduced
into this model.
Further, the sandpile evolution was identified
with the branching process and
the kinetic equations for the sand transfer
were written as equations of
chemical reactions of a special type. This
finer consideration allows one to
take into account more processes of toppling
inside a cell. In addition,
the generating function method proposed in the
same work makes all calculations
much simpler. Thus, the exponent $\tau $ was
defined with higher accuracy,
and different height site concentrations were
calculated.

Due to the fixed point independence from
the smallest scale dynamics, all the
results are valid for the model with an arbitrary rule of
toppling. Therefore,
this is correct for the Abelian sandpile model (ASM),
which was shown to be
exactly soluble by Dhar \cite{DD}. In fact, the values
of fixed point
concentrations are, indeed, in good agreement with the
exact values
calculated by Priezzhev \cite{VBP}, and value $\tau $
obtained in \cite{PVZ}
and \cite{EVI} is very close to $\tau =1.25$
proposed
by Priezzhev
{\it et al} in \cite{PKI}.

The relationship between the configurations
of the ASM and
spanning trees was found in \cite{MD1} for the
first time. Later the
one-to-one correspondence between the allowed ASM
configurations and the set
of spanning trees was shown in \cite{MD2}.
In the letter \cite{EVI},one interesting fact has
also been
noted: the fixed point probabilities of the sand being
transferred from the toppling cell to $one,two,three$ or
$four$ neighboring
cells were found to be almost equal to the
"branching probabilities of
spanning trees" which are the probabilities of a
random site of the
spanning tree having a coordination number $1,2,3$
or $4$, exactly
calculated in \cite{MDM}.It makes sense to verify
the hypothesis
about the equivalence of these probabilities and
in this way
to clarify a possible relation
between the ${\rm RG}$ approach and the spanning
tree statistics.

Excellent results of the RG approach on the square
lattice call for obtaining
the same results on other lattices. In our paper, we
consider the ${\rm RG}$ approach for the triangular
lattice. We
compare the obtained equilibrium concentrations
with the computer
simulations and the exponent $\tau $ of the avalanche
size distribution with its
exact value. We also calculate the branching
probabilities of the spanning
trees on the triangular lattice and compare them
with the fixed point
probabilities of the sand transfer into neighboring
sites. It is shown
that the hypothesis
about their coincidence at least does not have a
universal character.

\section{Renormalization approach.}
Let us consider a sandpile on the triangular lattice.
The height of each site
can take an integer value $Z=1,2,...~$ . If the
height of any site exceeds
the critical value $Z_c=6$, it topples.
At this moment, the particles can be transferred to
one or
several neighboring sites, increasing their heights
by one.
This toppling rule is
characterized by the vector ${{{{\bf p}}}}
=(p_1,p_2,...,p_6)$, where $p_i$ is the
probability for the site to topple
into $i$ neighboring cells.

Once we want to bring the consideration of this
model from the smallest
scale to a large one, we should define the successive
scale increasing
procedure. The step of this procedure consists in the
transition from the
sublattice $L_b$, which is constructed of the cells of
size $b$ on the
initial lattice $L$, to the sublattice $L_{b \sqrt{3}}$
with the cells of
size $b \sqrt{3}$. To realize it,
three cells of the sublattice $L_b$ are
unified into the block. This block is replaced by the
cell of $L_{b \sqrt{3}}$, as is shown in Fig.1a. All
bonds of $L_b$ outgoing from the given block
and coming into another block
are included into one renormalized
bond of $L_{b \sqrt{3}}$, as is shown in Fig 1b.

Now, we must describe the coarse grained sandpile
dynamics. In the following
calculations we use the $\rm RG$ approach in the
form proposed in \cite{PVZ}
and \cite{EVI}.
Let us define the {\it coarse grained variables}
for an arbitrary scale. They
are introduced in such a way that one can treat
the sandpile properties in
any scale like ones of the original sandpile.
Each cell of $L_b$ is
characterized by two vectors. The first one
is ${{{{\bf n}}}^{(b)}}%
=(n_A,n_B,n_C,n_D,n_E,n_F)$. This is the set
of probabilities for the $L_b$ cell
to behave, with respect to transformations
initiated by the dropped into
the cell particle, like a site on the initial
lattice with height $1,2,3,4,5$ or $6$,
respectively. In particular,
$n_F$ is the probability that
the cell is critical. The second vector,
${{\bf p}^{(b)}}$ is the analogue
of the vector ${\bf p}$ in the initial sandpile.

Let us represent the coarse grained dynamics of the
sandpile as the
following branching process.
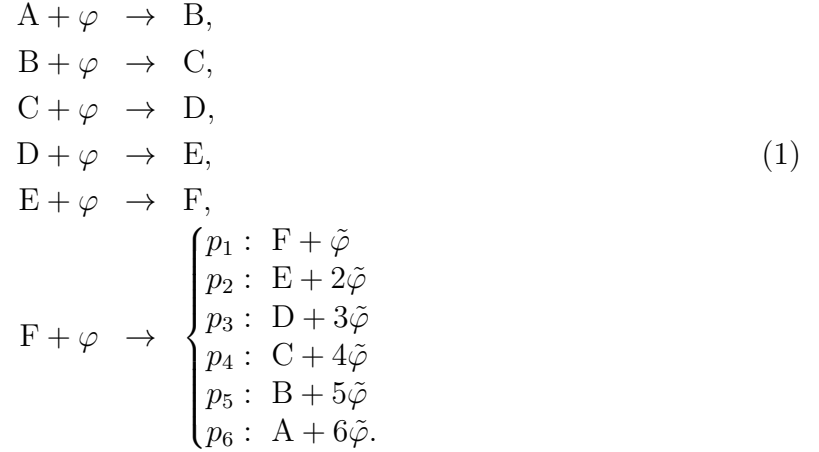
\begin{eqnarray}
{\rm A} + \varphi & \rightarrow & {\rm B}, \nonumber\\
{\rm B} + \varphi & \rightarrow & {\rm C}, \nonumber\\
{\rm C} + \varphi & \rightarrow & {\rm D}, \nonumber\\
{\rm D} + \varphi & \rightarrow & {\rm E}, \\
{\rm E} + \varphi & \rightarrow & {\rm F}, \nonumber\\
{\rm F} + \varphi & \rightarrow &
\left\{
\begin{minipage}{2.3cm}
$p_1:~ {\rm F} +  \tilde{\varphi}$\\
$p_2:~ {\rm E} + 2\tilde{\varphi}$\\
$p_3:~ {\rm D} + 3\tilde{\varphi}$\\
$p_4:~ {\rm C} + 4\tilde{\varphi}$\\
$p_5:~ {\rm B} + 5\tilde{\varphi}$\\
$p_6:~ {\rm A} + 6\tilde{\varphi}.$
\end{minipage}
\right.  \nonumber
\end{eqnarray}
Here the species $A,B,C,D,E,F$ are the cells
which behave like the sites
with heights $1,2,3,4,5,6$ , and $\varphi,
\tilde{\varphi}$ are the particles
obtained by the cell and transferred into other
cells, respectively.
This branching process can be reinterpreted
as irreversible chemical reactions for which
we can
write the kinetic
equations of the transport of $"sand"$ :
\begin{eqnarray}
\dot{n}_{\rm A} &=& n_\varphi~ (p_6~ n_{\rm F} -
n_{\rm A})\nonumber\\
\dot{n}_{\rm B} &=& n_\varphi~ (p_5~ n_{\rm F} +
n_{\rm A} - n_{\rm B})\nonumber\\
\dot{n}_{\rm C} &=& n_\varphi~ (p_4~ n_{\rm F} +
n_{\rm B} - n_{\rm C})\nonumber\\
\dot{n}_{\rm D} &=& n_\varphi~ (p_3~ n_{\rm F} +
n_{\rm C} - n_{\rm D})\\
\dot{n}_{\rm E} &=& n_\varphi~ (p_2~ n_{\rm F} +
n_{\rm D} - n_{\rm E})\nonumber\\
\dot{n}_{\rm D} &=& n_\varphi~ (p_1~ n_{\rm F} +
n_{\rm E} - n_{\rm F})\nonumber\\
\dot{n}_\varphi &=& n_\varphi~ (\bar{p}~n_{\rm F} -
1) + \bar{p}~ \nu \nabla^2 (n_\varphi n_{\rm F}) +
\eta ({\bf r}, t) \nonumber\\
{\bar p} &=& 6{p}_6^*+5{p}_5^*+4{p}_4^*+3{p}_3^*+
2{p}_2^*+{p}_1^* \nonumber
\end{eqnarray}
Here $n_\varphi$ denotes the concentrations of
particles $\varphi$,
$\eta({\bf r},t)$ is the noise term corresponding
to the randomly dropping
particles, $\nu$ is the diffusion coefficient ,which
is equal to $\frac16$
for the triangle lattice, $\bar p$ is the average
number of particles
outgoing from the toppling cell.

Performing the renormalization procedure we express
the parameters of
$L_{b\sqrt{3}}$ through ones of $L_b$.
\begin{equation}
{{{{{\bf n}}}^{(b\sqrt{3})}}=
{{{\bf n}}~}({{{\bf n}}}^{(b)},
{{{{\bf p}}}}^{(b)}),~~~~
{{{{{\bf p}}}}^{(b\sqrt{3})}}=
{{{{\bf p}}}~}({{{\bf n}}}^{(b)},
{{{{\bf p}}}}^{(b)})}
\end{equation}
The critical values of ${{{{\bf p}}}}^b$ and
${{{\bf n}}}^b$ obtained by taking the
limit $b\rightarrow \infty $ satisfy the
renormalization equations
\begin{equation}
{{{{{\bf n}}}^{*}}={{{\bf n}}~}({{{\bf n}}}^{*},
{{{{\bf p}}}}^{*}),~~~~
{{{{{\bf p}}}}^{*}}={{{{\bf p}}}~}({{{\bf n}}}^{*},
{{{{\bf p}}}}^{*})}
\label{REQ}
\end{equation}
Due to nonlocal properties of the sandpile dynamics,
writing down the first renormalization equation
does not seem
possible. To avoid it, we can exclude ${{{\bf n}}^*}$
from the second one.
We use equations (2) and the requirement for the
SOC state to be stationary. Taking
\begin{equation}
\dot {\bf n}^{*}=0,
\end{equation}
we can directly obtain
\begin{eqnarray}
{{\rm n}_{\rm A}^*} &=& \frac{{p}_6^*}
{\bar {p}}\nonumber \\
{\rm n}_{\rm B}^* &=& \frac{{p}_6^*+{p}_5^*}
{\bar p} \nonumber \\
{\rm n}_{\rm C}^* &=& \frac{{p}_6^*+{p}_5^*+
{p}_4^*}{\bar p} \nonumber \\
{\rm n}_{\rm D}^* &=& \frac{{p}_6^*+
{p}_5^*+{p}_4^*+{p}_3^*}{\bar p}           \\
{\rm n}_{\rm E}^* &=& \frac{{p}_6^*+{p}_5^*+{p}_4^*+
{p}_3^*+{p}_2^*}{\bar p} \nonumber \\
{\rm n}_{\rm F}^* &=& \frac{{p}_6^*+{p}_5^*+{p}_4^*+
{p}_3^*+{p}_2^*+{p}_1^*}{\bar p}~=~\frac{1}{\bar p}.
\nonumber
\end{eqnarray}
Substituting this into the second equation (\ref{REQ}) we obtain
a closed renormalization equation for the vector {\bf p}.
Search for the relationship between
${{{{\bf p}}}}^{(b)}$ and
${{{{\bf p}}}}^{(b \sqrt{3})}$ is reduced to
counting the processes at the scale $b$ leading to the
toppling of the cell at the scale  ${b \sqrt{3}}$.
To this end, we,
following \cite{EVI}, introduce the generating function of
the toppling of the cell at an arbitrary scale.
If we denote $six$ directions outgoing from the cell
by $x_1,...,x_6$, the generating function of toppling
of the sublattice $L^b$ cell will be given by the formula
\begin{eqnarray}
{{\rm \sigma}(x_1,x_2,x_3,x_4,x_5,x_6)=\frac{p_1^{(b)}}6
(x_1+x_2+x_3+x_4+x_5+x_6)}+\nonumber \\
{\frac{{p_2}^{(b)}}{15}(x_1 x_2+x_1 x_3+x_1 x_4+ ... +x_5
x_6)+\frac{{p_3}^{(b)}}{20}{\rm \sigma}'''+~~~~~~~~ }\\
\frac{{p_4}^{(b)}}{15}{\rm \sigma}''''+
\frac{{p_5}^{(b)}}{6}{\rm \sigma}'''''+
{{p_6}^{(b)}(x_1 x_2 x_3 x_4 x_5 x_6),
~~~~~~~~~~~~~~~}\nonumber
\end{eqnarray}
where ${\rm \sigma}$ with $i$ primes is the sum of
all possible products $x_{k_1} x_{k_2}...x_{k_i}$
of $i$ different
terms $x_{k_1},...,x_{k_i}$ .
This function counts all processes leading to sand
transition from the cell
of $L_b$.
The coefficient of each polynomial term
is the probability for
particles to go to the corresponding
directions
after the cell toppling.
This function has the following
properties:

a)~~ If the arguments corresponding to
any directions
are replaced by zero, the function
counts the toppling
processes that don't send the particles
to these directions.

b)~~ If they
are replaced by unit, the function
counts the toppling
processes irrespective of outgoing to
these directions.

c)~~If we want to describe the toppling of
a couple of cells,
where the particle sent by the first cell
induces the toppling
of the second one, we must replace the argument
labeling the
direction of outgoing of this particle by the
$\sigma$-function
of toppling of the second cell.This function contains also
the processes in which the particle does not go
in the given direction
and the second cell does not topple.
It is necessary to subtract them,
if we want to take into account only processes
spreading to
both cells.

d)~~In addition, it is easy to see that
the function is normalized
so that $\sigma(1,...,1)=1$.

Combining $\sigma$-functions in different ways
we can build the functions
that count the specific topplings.
Once we manage to link the generating
functions at different scales, we express the
probabilities ${{{\bf p}}}$
at a given scale through ones at a new scale.
To this end, we write the generating function for the
block consisting of
three cells of $L_b$.
 This function must count the processes leading to the
transition of
particles from this block to some neighboring blocks.
We can perform this using $\sigma$-function
properties  outlined above.
In addition, according to
\cite{PVZ} we must consider only processes which
match the spanning condition, i.e.
with the characteristic size equal to the size of a cell.
In our case, these are
relaxation processes in which more than one site
of $L_{b}$ participates. All
three types of blocks, whose topplings match the
spanning condition
and their relaxation schemes are shown in Fig 2.
The rest of the
schemes can be obtained from the shown ones by
rotations.

For going over to the sublattice $L_{b\sqrt{3}}$,
we replace the block considered by
the cell of $L_{b\sqrt{3}}$, respectively
renaming the outgoing directions,
as is shown in Fig. 1b.
To give the obtained function the meaning of
the generating function, we must
subtract all processes not going out of the block
and  normalize the result.
Thus, we obtain
the generating functions ${\rm
\Sigma}_j(y_1,y_2,y_3,y_4,y_5,y_6)(j=a,b,c)$
for the topplings of blocks of three
kinds: $a,b$ and $c$.
For example, the generating function
for relaxation of the block of type $a$
is the following:
\begin{eqnarray}
{\rm \Sigma}_a(y_1,y_2,y_3,y_4,y_5,y_6)~=\frac
{\tilde{\rm \Sigma}_a(y_1,...,y_6)-
\tilde{\rm \Sigma}_a(0,...,0)}
{\tilde{\rm \Sigma}_a(1,...,1)-
\tilde{\rm \Sigma}_a(0,...,0)}
\end{eqnarray}
where
\begin{eqnarray}
{\tilde{\rm \Sigma}_a(y_1,y_2,y_3,y_4,y_5,y_6)~=
~~~~~~~~~~~~~~~~~~~~~~~~~~~~~~~~~~~~~~~~~~~~~~
~~~~~~~~~~~~~~~~~} \nonumber \\
{({\rm \sigma}(y_1,{\rm \sigma}(y_2,y_2,y_3,1,1,y_1)
,1,y_5,y_6,y_6)-
{\rm \sigma}(y_1,0,1,y_5,y_6,y_6)+~~~~~~~~~~~~~~~~}\\
{{\rm \sigma}(y_2,y_2,y_3,1,{\rm
\sigma}(y_1,1,1,y_5,y_6,y_6),y_1)-
{\rm \sigma}(y_2,y_2,y_3,1,0,y_1))+
c.p.~~~~~~~~~~}\nonumber
\end{eqnarray}
The arguments $y_1~...~y_6$ denote the directions on the
lattice at the new scale.
The terms $c.p.$ are obtained from the first
term by the cyclic permutations
of arguments corresponding to the rotation
by the angles ${\rm \frac{2 \pi}{3}}$
and ${\rm \frac{4 \pi}3}$.
The generating functions for the blocks of other types
look analogously
but they are too cumbersome to be given here.
To write a general generating function, we
accomplish summation over all types of blocks.
Different types of blocks appear with the
different probabilities
which depend only on their heights.
Therefore, we must multiply their generating functions
by the corresponding weights.
After summation the result must again be normalized
\begin{eqnarray}
{{\rm \Sigma}(y_1,y_2,y_3,y_4,y_5,y_6)~=~\frac{
{\rm \tilde{\Sigma}}
(y_1,...,y_6)}{{\rm \tilde{\Sigma}}(1,...,1)}~~~~~~~~~~
~~~~~~~~~~~~}\nonumber\\
\nonumber \\
{{\rm \tilde{\Sigma}}={\rm w_a}{\rm \tilde{\Sigma}}_a +
{\rm w_b}{\rm \tilde{\Sigma}}_b +
{\rm w_c}{\rm \tilde{\Sigma}}_c~~~~~~~~~~~~~~~~~~~
~~~~~~~~~~~~~~~~~~~~~~~~~~~~~~}\\
\nonumber \\
{{\rm w}_a=3{{n_{\rm F}}^2}(n_{\rm A}+n_{\rm
B}+n_{\rm C}+n_{\rm D}),~~~~
{\rm w}_b=3{{n_{\rm F}}^2}n_{\rm E},~~~
{\rm w}_c={n_{\rm F}}^3~~~~}.\nonumber
\end{eqnarray}
Thus, we have the normalized generating function
describing the toppling of a cell of the
lattice
$L_{b \sqrt{3}}$. According to the RG ideology, all
particles outgoing from a
cell in some direction become one renormalized
particle. Therefore, the second
powers of $y_i$ must be changed to the first powers.
Now the coefficient of each
term is again the probability of the respective
process of toppling of the size
$b \sqrt{3}$ cell. Taking the coefficients of
sums analogous to (1), we obtain
new probabilities ${{{{\bf p}}}}^{(b \sqrt{3})}$
expressed through ${{{{\bf p}}}}^{(b)}$ and
${{{\bf n}}}^{(b)}$.
Eventually, we have sought a system of renormalization
equations.
Solving this system, we obtain the fixed point
values of ${{{{\bf p}}}}$ and
${{{\bf n}}}$.They are shown in tables 1 and 2.

According to Pietronero {\it et al}, the critical
exponent of the avalanche distribution $\tau $
may be expressed through
the critical point parameters as follows.
The probability of an avalanche of the
linear size $r$ is $P(r)\,dr=r^{(1-2\tau )}\,dr$.
Thus, the probability for
the avalanche to cover the cell of the size $b$
dying beyond it is given by the
formula:
\begin{equation}
K=\frac{\int_b^{b\sqrt{3}}{P}(r)\,dr}{\int_b^\infty
{P}(r)\,dr}=1-(\sqrt{3})^{2(1-\tau )}
\end{equation}
On the other hand, $K$ can be expressed through the
fixed point parameters
\begin{equation}
K=p_1^{*}(1-n_F)+p_2^{*}(1-n_F^{*})^2+...
+p_6^{*}(1-n_F^{*})^6
\end{equation}
Using these expressions we obtain
\begin{equation}
\tau =1-\frac{\ln (1-K)}{\ln \,3}=1.366
\end{equation}

\section{Results and discussion.}
Obtaining the height probabilities
in the sandpile on the triangular
lattice by the RG approach seems to be
efficient because of the absence of an
exact result. While the RG
approach is relatively simple, the exact
calculations require applying a very
complex technique even in the simplest
case. In Table 1, we compare the
fixed point height probabilities with the
results of numerical simulation .One can
notice a very similar behavior of the
numerical and RG results.
\begin{center}
\begin{tabular}{lcccccc}
\hline\hline
\vspace{-4mm}\\
 & ~~~$n_1^*~~~$ & ~~~$n_2^*$~~~ & ~~~$n_3^*$~~~ & ~~~$n_4^*$~~~ & ~~~$n_5^*$~~~ & ~~~$n_6^*$~~~ \\
\vspace{-4mm}\\
\hline
\vspace{-3mm}\\
RG & 0.036 & 0.135 & 0.197 & 0.209 & 0.210 & 0.210  \\
\vspace{-3mm}\\
Numerical & 0.054 & 0.092 & 0.139 & 0.188 & 0.240 & 0.281\\
\vspace{-3mm}\\
\hline \hline \\
\end{tabular}

{\sc Table 1.} -- {\sc \ }{\small Fixed point
height probabilities in
comparison with the numerical results.\\
The numerical simulations were performed on the
lattice $300\times 300$ with
statistics $10^6$ configurations.}
\end{center}

Some distinction
of their values is stipulated
by initial roughness of the RG approach
concerned with discrete increasing of
the scale and averaged consideration of
the coarse grained dynamics. However, the
advantage of this RG scheme is a quite
attractive character of the obtained
fixed point.

The obtained value $\tau =1.366$ is close to
the exact one $1.25$ although the
coincidence is rather worse than in
the square lattice.
This can be explained 
by a small size of a renormalized cell.
The renormalization scheme thus
constructed takes into account
only a minimal number of toppling
processes. Therefore, these calculations don't
give high accuracy.
However, this result allows us to speak about the
stability for a given RG.

At the end, in Table 2, we compare the fixed
point probabilities
of the sand transfer $p_k$ with the branching
probabilities of spanning
trees calculated in Appendix.
\begin{center}
\begin{tabular}{lcccccc}
\hline\hline
\vspace{-4mm}\\
 & ~~~$p_1^*~~~$ & ~~~$p_2^*$~~~ & ~~~$p_3^*$~~~ & ~~~$p_4^*$~~~ & ~~~$p_5^*$~~~ & ~~~$p_6^*$~~~ \\
\vspace{-4mm}\\
\hline
\vspace{-3mm}\\
RG & 0.0000179 & 0.00226 & 0.0558 & 0.296 & 0.471 & 0.174  \\
\vspace{-3mm}\\
{\small Branching}&                                                \\
{\small probabilities} & 0.322 & 0.417 & 0.207 & 0.0488 & 0.00553 & 0.000241\\
\hline \hline\\
\end{tabular}

{\sc Table 2.} -- {\small Fixed point
probabilities $p_k$ of the sand being
transported to k neighboring cells and
branching probabilities of spanning
trees on the triangular lattice.\\}
\end{center}

The comparison of the probabilities of the 
sand transfer with the
branching probabilities of 
spanning trees is very surprizing.
As these 
quantities for the square 
lattice are very close, a similar behavior can
be expected in our case as well.
However, one can see quite an
inverse order of the obtained values
for the example considered.

Thus, we can conclude that the hypothesis
about their coincidence is not confirmed
in the case of triangular lattice.
Further clarification of this question
seems to be related to more
complete investigations of a sandpile
on a square lattice.
\section*{Acknowledgments}
We are grateful to Prof. V.B.~Priezzhev
and E.V.~Ivashkevich  for fruitful discussions and
critical reading of the manuscript. We also want to thank
D.G.~Gaidashev for useful remarks.
\begin{appendix}
\section*{Appendix}
The branching probability of a spanning tree
$p_k$ is the probability for any site of a
random spanning tree to have a
coordination number $k$. The
effective tool used for evaluating branching
probabilities is provided by the Kirchhoff theorem
about spanning trees. The
method of applying it is not novel.
Below we follow the methods outlined in
the article \cite{VBP}. Here we recall
its main features. To formulate the
Kirchhoff theorem we consider the
arbitrarily connected graph ${\bf L}$
consisting of $n+1$ sites. Let us fix
any site which will be a root.
The weight $x_{ij}$ corresponds to
each bond of ${\bf L}$
edged by the sites $i$ and $j$.
Let $\Delta(x)$ be a $n \times n$ matrix:
\begin{equation}
\Delta_{ij}(x)= \left\{
\begin{array}{ccc}
\sum_{k} x_{ik} & , & i=j \\
{-x_{ij}} & , & i
{\rm ~and~} j {\rm ~are~adjacent~sites} \\ 0 & ,
& {\rm otherwise}.
\end{array}
\right. 
\end{equation}
While the summation is performed over all
sites of $G$ $i$ and $j$ run
over all sites different from the root.
Then, according to the Kirchhoff
theorem,
\begin{equation}
g(x)=det\Delta(x)
\end{equation}
is nothing but a generating function of the rooted
spanning trees on this
graph. If all $x_{ij}$ are equal to $1$, the matrix
\begin{equation}
\Delta=\Delta(x_{ij}=1)
\end{equation}
is the discrete Laplacian on graph ${\bf L}$ and
\begin{equation}
N=det\Delta
\end{equation}
gives a total number of the spanning trees on
this graph.

For calculating a number of spanning trees on the
triangular lattice $L$ whose
random site $i$ has $k$ adjacent sites,
let us consider the
lattices $L^{\prime}$ with the defect in the
cell containing site $i$.
In other words, we deal with lattices obtained
from $L$ by deleting one
or several bonds outgoing from $i$. If we
leave only one bond connected with
$i$, all trees on this new lattice will
occupy it. Summing over all possible
directions of this bond, we obtain the
number of spanning trees the site $i$
of which has only one adjacent site.
In the same manner we can link the numbers of
spanning trees with other coordination
numbers of site $i$ with total
numbers of spanning trees on the
lattices with a local defect

\begin{equation}
N_i^{\prime }=\sum\limits_{k=1}^i
\frac{C_6^iC_i^k}{C_6^k}N_k
\end{equation}

where $N_k$ is the number of spanning trees in which
the site $i$ has the
coordination number $k$. $N_k^{\prime}$ is the
sum of total numbers of
spanning trees on the lattices obtained from $L$ by
deleting all bonds
connected with site $i$ except $k$ ones which is taken
over all possible positions of
these bonds. Dividing $N_k$ by the total number
of spanning trees, we obtain
the branching probabilities.

Thus, we reduce the problem of searching for
branching probabilities search to the
calculation of the numbers of spanning trees on the
lattices with the defect.
To this end, we use the Kirchhoff theorem.
According to it, the number of spanning
trees on the lattice with a defect is given by
the formula
\begin{equation}
N^{\prime}=det\Delta^{\prime}
\end{equation}
$\Delta^{\prime}$ is the discrete Laplacian on a
new lattice. If we refer to the
difference between $\Delta$ and
$\Delta^{\prime}$ as a defect matrix $\delta$,
branching probabilities are expressed
through the terms of the following form:
\begin{equation}
\frac{N^{\prime}}N=\frac{det
\Delta^{\prime}}{det\Delta}=det(I+G\delta)
\end{equation}
$I$ is the unit matrix; $G$ satisfies the matrix equation
\begin{equation}
{\Delta}G=I
\end{equation}
In the thermodynamic limit $G$ is nothing but the
Green function of the Laplace
equation on the triangle lattice. For an infinite
lattice $G$ depends only on
the difference of site coordinates rather
than their values
\begin{equation}
G({\bf r_1-r_2})=\int^{2\pi}_0 \int^{2\pi}_0
\frac{\cos(\alpha(x_1-x_2)+\beta(y_1-y_2))-1}
{1-\frac{1}{3}(\cos{\alpha}+cos{\beta}%
+cos(\alpha+\beta))}\,\frac{d\alpha}{2\pi}\,
\frac{d\beta}{2\pi}
\end{equation}
Due to defect localization, the matrix $\delta$
has only a finite number of
nonzero elements. Thus, we should calculate
the $7\times7$ matrix
determinants. In this way, we obtain
the following results:
\begin{eqnarray}
{{\it p_1}= {-{\frac{{25}}{{108}}} +
{\frac{{324\,{\sqrt{3}}}}{{{{\pi }^5}}}}
- {\frac{{540}}{{{{\pi }^4}}}} +
{\frac{{66\,{\sqrt{3}}}}{{{{\pi }^3}}}} + {%
\frac{{14}}{{{{\pi }^2}}}} -
{\frac{{55}}{{12\,{\sqrt{3}}\,\pi }}}}} \nonumber \\ {%
{\it p_2}= {-{\frac{{235}}{{144}}} -
{\frac{{1620\,{\sqrt{3}}}}{{{{\pi }^5}}}}
+ {\frac{{3213}}{{{{\pi }^4}}}} -
{\frac{{702\,{\sqrt{3}}}}{{{{\pi }^3}}}} +
{\frac{{277}}{{2\,{{\pi }^2}}}} +
{\frac{{223}}{{12\,{\sqrt{3}}\,\pi }}}}}  \nonumber
\\ {{\it p_3}= {{\frac{{17}}{{36}}} +
{\frac{{3240\,{\sqrt{3}}}}{{{{\pi }^5}}}%
} - {\frac{{7452}}{{{{\pi }^4}}}} +
{\frac{{2160\,{\sqrt{3}}}}{{{{\pi }^3}}}}
- {\frac{{838}}{{{{\pi }^2}}}} +
{\frac{{241}}{{2\,{\sqrt{3}}\,\pi }}}}} \\{%
{\it p_4}= {{\frac{{2489}}{{216}}} -
{\frac{{3240\,{\sqrt{3}}}}{{{{\pi }^5}}}}
+ {\frac{{8478}}{{{{\pi }^4}}}} -
{\frac{{2928\,{\sqrt{3}}}}{{{{\pi }^3}}}}
+ {\frac{{1489}}{{{{\pi }^2}}}} -
{\frac{{2203}}{{6\,{\sqrt{3}}\,\pi }}}}} \nonumber
\\ {{\it p_5}= {-{\frac{{115}}{9}} +
{\frac{{1620\,{\sqrt{3}}}}{{{{\pi }^5}}}}
- {\frac{{4752}}{{{{\pi }^4}}}} +
{\frac{{1854\,{\sqrt{3}}}}{{{{\pi }^3}}}}
- {\frac{{1088}}{{{{\pi }^2}}}} +
{\frac{{3857}}{{12\,{\sqrt{3}}\,\pi }}}}}\nonumber
\\ {{\it p_6}= {{\frac{{175}}{{48}}} -
{\frac{{324\,{\sqrt{3}}}}{{{{\pi }^5}}}%
} + {\frac{{1053}}{{{{\pi }^4}}}} -
{\frac{{450\,{\sqrt{3}}}}{{{{\pi }^3}}}}
+ {\frac{{569}}{{2\,{{\pi }^2}}}} -
{\frac{{355}}{{4\,{\sqrt{3}}\,\pi }}}}}
\nonumber \\ \nonumber
\end{eqnarray}
\end{appendix}

\newpage
{\bf Figure Captions}
\begin{description}

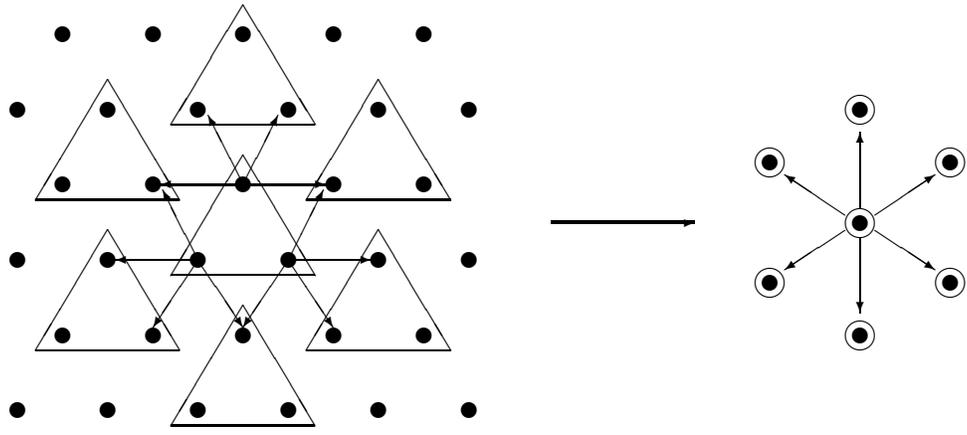
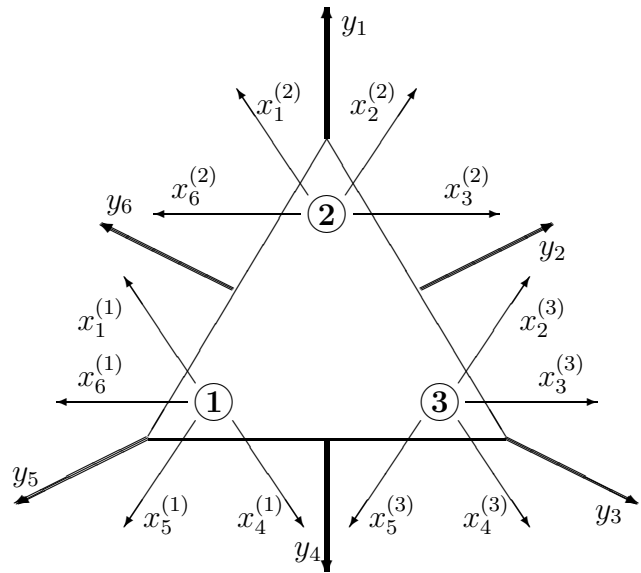
\item[Fig.1:]

~a)~~Transformation from the lattice $L_b$
to $L_{b \sqrt{3}}$.
It is performed in such a way that the new
lattice is again triangular.

~~~b)~~Block of three cells of the sublattice $L_b$.
This block becomes a cell on the new lattice.
We show that the directions
outgoing from the cells, forming the
block, are coupled to
the directions of the lattice at the next scale.
For example: $x_1^{(2)},x_2^{(2)}$ are replaced by $y_1$
and $x_3^{(2)},x_2^{(3)}$ are replaced by $y_2$.
The superscript denotes
the cell from which the given direction outgoes.

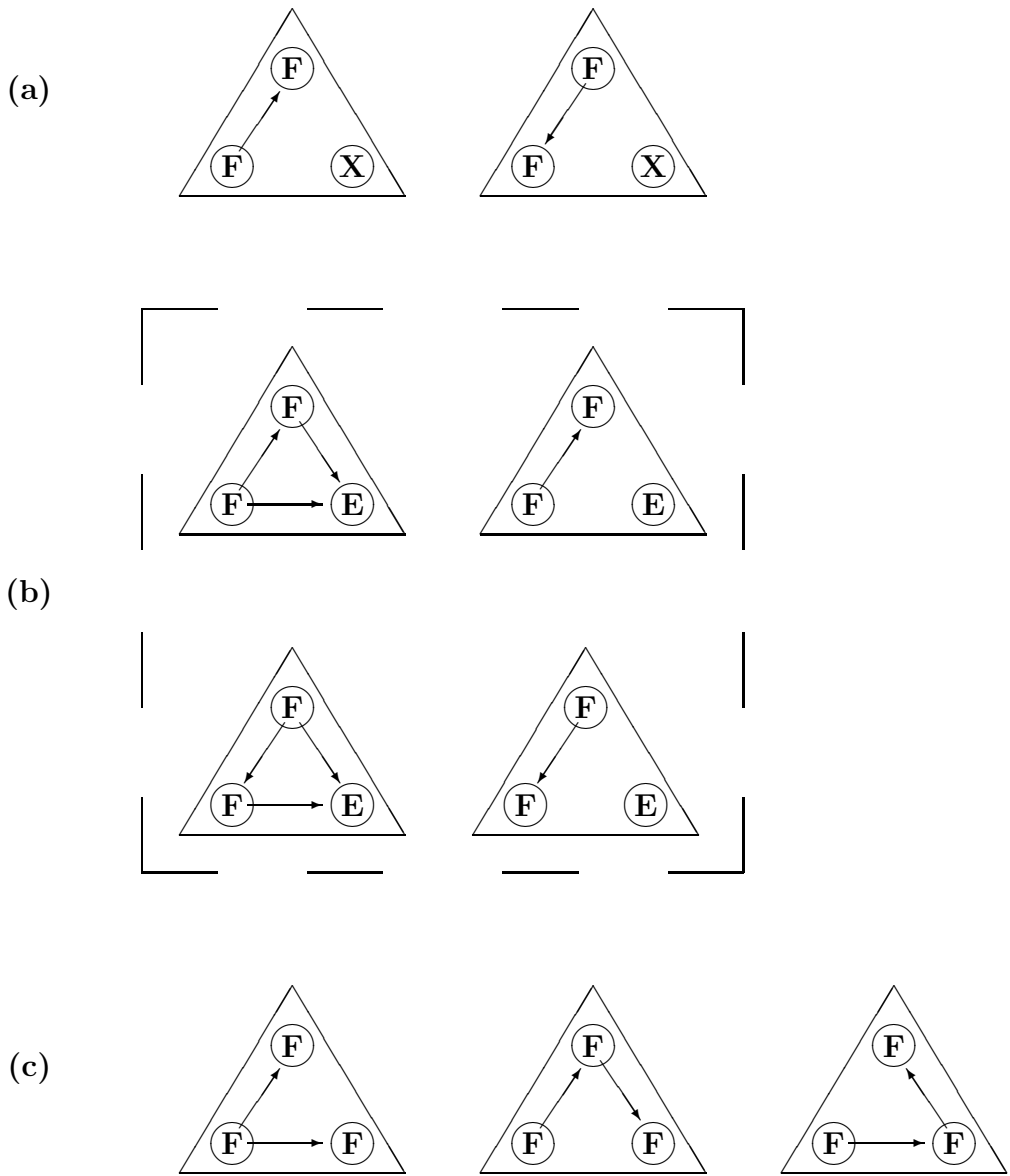
\item[Fig.2:] Three types of blocks, whose
toppling should be taken
into account, and their relaxation schemes.
The other blocks can be obtained
from these ones by rotations.
The letters in the circles denote
the height of the cells.
X can take the value A,B,C or D.
\end{description}

\bf
\begin{figure}[p]
\unitlength=1.00mm
\special{em:linewidth 0.4pt}
\linethickness{0.4pt}
\begin{picture}(141.00,154.00)
\put(21.00,150.00){\circle*{2.00}}
\put(33.00,150.00){\circle*{2.00}}
\put(27.00,140.00){\circle*{2.00}}
\put(39.00,140.00){\circle*{2.00}}
\put(21.00,130.00){\circle*{2.00}}
\put(33.00,130.00){\circle*{2.00}}
\put(27.00,120.00){\circle*{2.00}}
\put(39.00,120.00){\circle*{2.00}}
\put(45.00,130.00){\circle*{2.00}}
\put(51.00,120.00){\circle*{2.00}}
\put(45.00,150.00){\circle*{2.00}}
\put(57.00,150.00){\circle*{2.00}}
\put(51.00,140.00){\circle*{2.00}}
\put(63.00,140.00){\circle*{2.00}}
\put(57.00,130.00){\circle*{2.00}}
\put(63.00,120.00){\circle*{2.00}}
\put(69.00,130.00){\circle*{2.00}}
\put(75.00,120.00){\circle*{2.00}}
\put(69.00,150.00){\circle*{2.00}}
\put(75.00,140.00){\circle*{2.00}}
\put(21.00,110.00){\circle*{2.00}}
\put(33.00,110.00){\circle*{2.00}}
\put(45.00,110.00){\circle*{2.00}}
\put(57.00,110.00){\circle*{2.00}}
\put(69.00,110.00){\circle*{2.00}}
\put(115.00,133.00){\circle*{2.00}}
\put(139.00,117.00){\circle*{2.00}}
\put(127.00,125.00){\circle*{2.00}}
\put(127.00,110.00){\circle*{2.00}}
\put(115.00,117.00){\circle*{2.00}}
\put(139.00,133.00){\circle*{2.00}}
\put(27.00,100.00){\circle*{2.00}}
\put(39.00,100.00){\circle*{2.00}}
\put(51.00,100.00){\circle*{2.00}}
\put(63.00,100.00){\circle*{2.00}}
\put(75.00,100.00){\circle*{2.00}}
\put(127.00,125.00){\circle{4.00}}
\put(127.00,140.00){\circle{4.00}}
\put(139.00,133.00){\circle{4.00}}
\put(139.00,117.00){\circle{4.00}}
\put(127.00,110.00){\circle{4.00}}
\put(115.00,117.00){\circle{4.00}}
\put(115.00,133.00){\circle{4.00}}
\put(127.00,140.00){\circle*{2.00}}
\put(125.00,126.00){\vector(-3,2){8.00}}
\put(45.00,130.00){\vector(1,0){11.00}}
\put(45.00,130.00){\vector(-1,0){11.00}}
\put(51.00,120.00){\vector(1,0){11.00}}
\put(39.00,120.00){\vector(-1,0){11.00}}
\put(39.00,120.00){\vector(-1,2){4.67}}
\put(39.00,120.00){\vector(2,-3){6.00}}
\put(51.00,120.00){\vector(-2,-3){6.00}}
\put(51.00,120.00){\vector(1,2){4.67}}
\put(51.00,120.00){\vector(2,-3){6.00}}
\put(45.00,130.00){\vector(-1,2){4.67}}
\put(45.00,130.00){\vector(1,2){4.67}}
\put(39.00,120.00){\vector(-2,-3){6.00}}
\put(125.00,124.00){\vector(-3,-2){8.00}}
\put(129.00,126.00){\vector(3,2){8.00}}
\put(129.00,124.00){\vector(3,-2){8.00}}
\put(127.00,127.00){\vector(0,1){10.00}}
\put(127.00,123.00){\vector(0,-1){10.00}}
\put(15.00,140.00){\circle*{2.00}}
\put(15.00,120.00){\circle*{2.00}}
\put(15.00,100.00){\circle*{2.00}}
\put(45.00,154.00){\line(-3,-5){9.62}}
\put(45.00,154.00){\line(3,-5){9.62}}
\put(54.50,138.00){\line(-1,0){19.10}}
\put(63.00,144.00){\line(-3,-5){9.62}}
\put(63.00,144.00){\line(3,-5){9.62}}
\put(72.50,128.00){\line(-1,0){19.10}}
\put(45.00,134.00){\line(-3,-5){9.62}}
\put(45.00,134.00){\line(3,-5){9.62}}
\put(54.50,118.00){\line(-1,0){19.10}}
\put(63.00,124.00){\line(-3,-5){9.62}}
\put(63.00,124.00){\line(3,-5){9.62}}
\put(72.50,108.00){\line(-1,0){19.10}}
\put(45.00,114.00){\line(-3,-5){9.61}}
\put(45.00,114.00){\line(3,-5){9.61}}
\put(54.50,98.00){\line(-1,0){19.10}}
\put(27.00,144.00){\line(-3,-5){9.62}}
\put(27.00,144.00){\line(3,-5){9.62}}
\put(36.50,128.00){\line(-1,0){19.10}}
\put(27.00,124.00){\line(-3,-5){9.62}}
\put(27.00,124.00){\line(3,-5){9.62}}
\put(36.50,108.00){\line(-1,0){19.10}}
\put(81.00,85.00){\makebox(0,0)[cc]{(a)}}
\special{em:linewidth 1.pt}
\linethickness{1.pt}
\put(86.00,125.00){\vector(1,0){19.00}}
\unitlength=2.50mm
\special{em:linewidth 0.4pt}
\linethickness{0.4pt}
\put(32.50,19.00){\line(-3,-5){9.50}}
\put(32.50,19.00){\line(3,-5){9.50}}
\put(42.00,3.00){\line(-1,0){19.00}}
\put(25.50,16.50){\makebox(0,0)[cc]{$x_6^{(2)}$}}
\put(30.00,21.00){\makebox(0,0)[cc]{$x_1^{(2)}$}}
\put(35.00,21.00){\makebox(0,0)[cc]{$x_2^{(2)}$}}
\put(40.00,16.50){\makebox(0,0)[cc]{$x_3^{(2)}$}}
\put(45.00,6.50){\makebox(0,0)[cc]{$x_3^{(3)}$}}
\put(44.00,9.50){\makebox(0,0)[cc]{$x_2^{(3)}$}}
\put(41.00,-1.00){\makebox(0,0)[cc]{$x_4^{(3)}$}}
\put(36.00,-1.00){\makebox(0,0)[cc]{$x_5^{(3)}$}}
\put(29.00,-1.00){\makebox(0,0)[cc]{$x_4^{(1)}$}}
\put(24.00,-1.00){\makebox(0,0)[cc]{$x_5^{(1)}$}}
\put(20.50,6.50){\makebox(0,0)[cc]{$x_6^{(1)}$}}
\put(20.50,9.50){\makebox(0,0)[cc]{$x_1^{(1)}$}}
\put(34.00,25.00){\makebox(0,0)[cc]{$y_1$}}
\put(44.50,13.00){\makebox(0,0)[cc]{$y_2$}}
\put(47.50,-1.00){\makebox(0,0)[cc]{$y_3$}}
\put(31.50,-3.00){\makebox(0,0)[cc]{$y_4$}}
\put(21.50,15.50){\makebox(0,0)[cc]{$y_6$}}
\put(16.50,1.00){\makebox(0,0)[cc]{$y_5$}}

\put(32.40,19.00){\vector(0,1){7.00}}
\put(32.50,19.00){\vector(0,1){7.00}}
\put(32.60,19.00){\vector(0,1){7.00}}
\put(32.40,3.00){\vector(0,-1){7.00}}
\put(32.50,3.00){\vector(0,-1){7.00}}
\put(32.60,3.00){\vector(0,-1){7.00}}
\put(41.95,3.05){\vector(2,-1){7.00}}
\put(42.00,3.10){\vector(2,-1){7.00}}
\put(42.05,3.15){\vector(2,-1){7.00}}
\put(22.90,3.15){\vector(-2,-1){7.00}}
\put(23.00,3.10){\vector(-2,-1){7.00}}
\put(23.10,3.05){\vector(-2,-1){7.00}}
\put(37.45,11.05){\vector(2,1){7.00}}
\put(37.50,11.00){\vector(2,1){7.00}}
\put(37.55,10.95){\vector(2,1){7.00}}
\put(27.45,10.95){\vector(-2,1){7.00}}
\put(27.50,11.00){\vector(-2,1){7.00}}
\put(27.55,11.05){\vector(-2,1){7.00}}

\put(32.50,15.00){\circle{2.00}}
\put(32.50,15.00){\makebox(0,0)[cc]{2}}
\put(38.50,5.00){\circle{2.00}}
\put(38.50,5.00){\makebox(0,0)[cc]{3}}
\put(26.50,5.00){\circle{2.00}}
\put(26.50,5.00){\makebox(0,0)[cc]{1}}
\put(31.10,15.00){\vector(-1,0){7.80}}
\put(33.90,15.00){\vector(1,0){7.80}}
\put(39.90,5.00){\vector(1,0){7.00}}
\put(25.10,5.00){\vector(-1,0){7.00}}
\put(33.50,16.00){\vector(2,3){3.78}}
\put(31.50,16.00){\vector(-2,3){3.78}}
\put(39.50,6.00){\vector(2,3){3.78}}
\put(25.50,6.00){\vector(-2,3){3.78}}
\put(37.50,4.00){\vector(-2,-3){3.78}}
\put(27.50,4.00){\vector(2,-3){3.78}}
\put(39.50,4.00){\vector(2,-3){3.78}}
\put(25.50,4.00){\vector(-2,-3){3.78}}
\put(32.50,-8.00){\makebox(0,0)[cc]{(b)}}
\put(32.50,-12.00){\makebox(0,0)[cc]{Fig.1.}}
\end{picture}
\end{figure}
\begin{figure}[p]
\unitlength=1.00mm
\special{em:linewidth 0.4pt}
\linethickness{0.4pt}
\begin{picture}(140.00,145.00)
\put(38.00,124.00){\circle{6.00}}
\put(30.00,137.00){\circle{6.00}}
\put(22.00,124.00){\circle{6.00}}
\put(23.00,126.00){\vector(2,3){5.33}}
\put(30.00,137.00){\makebox(0,0)[cc]{F}}
\put(22.00,124.00){\makebox(0,0)[cc]{F}}
\put(38.00,124.00){\makebox(0,0)[cc]{X}}
\put(30.00,145.00){\line(3,-5){15.00}}
\put(45.00,120.00){\line(-1,0){30.00}}
\put(15.00,120.00){\line(3,5){15.00}}
\put(38.00,79.00){\circle{6.00}}
\put(30.00,92.00){\circle{6.00}}
\put(22.00,79.00){\circle{6.00}}
\put(23.00,81.00){\vector(2,3){5.33}}
\put(24.00,79.00){\vector(1,0){9.99}}
\put(30.00,92.00){\makebox(0,0)[cc]{F}}
\put(22.00,79.00){\makebox(0,0)[cc]{F}}
\put(38.00,79.00){\makebox(0,0)[cc]{E}}
\put(30.00,100.00){\line(3,-5){15.00}}
\put(15.00,75.00){\line(3,5){15.00}}
\put(78.00,79.00){\circle{6.00}}
\put(70.00,92.00){\circle{6.00}}
\put(62.00,79.00){\circle{6.00}}
\put(63.00,81.00){\vector(2,3){5.33}}
\put(70.00,92.00){\makebox(0,0)[cc]{F}}
\put(62.00,79.00){\makebox(0,0)[cc]{F}}
\put(78.00,79.00){\makebox(0,0)[cc]{E}}
\put(70.00,100.00){\line(3,-5){15.00}}
\put(85.00,75.00){\line(-1,0){30.00}}
\put(55.00,75.00){\line(3,5){15.00}}
\put(15.00,75.00){\line(1,0){30.00}}
\put(38.00,-6.00){\circle{6.00}}
\put(30.00,7.00){\circle{6.00}}
\put(22.00,-6.00){\circle{6.00}}
\put(23.00,-4.00){\vector(2,3){5.33}}
\put(24.00,-6.00){\vector(1,0){9.99}}
\put(30.00,7.00){\makebox(0,0)[cc]{F}}
\put(22.00,-6.00){\makebox(0,0)[cc]{F}}
\put(38.00,-6.00){\makebox(0,0)[cc]{F}}
\put(30.00,15.00){\line(3,-5){15.00}}
\put(15.00,-10.00){\line(3,5){15.00}}
\put(78.00,-6.00){\circle{6.00}}
\put(70.00,7.00){\circle{6.00}}
\put(62.00,-6.00){\circle{6.00}}
\put(63.00,-4.00){\vector(2,3){5.33}}
\put(71.00,5.00){\vector(2,-3){5.30}}
\put(70.00,7.00){\makebox(0,0)[cc]{F}}
\put(62.00,-6.00){\makebox(0,0)[cc]{F}}
\put(78.00,-6.00){\makebox(0,0)[cc]{F}}
\put(70.00,15.00){\line(3,-5){15.00}}
\put(85.00,-10.00){\line(-1,0){30.00}}
\put(55.00,-10.00){\line(3,5){15.00}}
\put(15.00,-10.00){\line(1,0){30.00}}
\put(118.00,-6.00){\circle{6.00}}
\put(110.00,7.00){\circle{6.00}}
\put(102.00,-6.00){\circle{6.00}}
\put(104.00,-6.00){\vector(1,0){9.99}}
\put(110.00,7.00){\makebox(0,0)[cc]{F}}
\put(102.00,-6.00){\makebox(0,0)[cc]{F}}
\put(118.00,-6.00){\makebox(0,0)[cc]{F}}
\put(110.00,15.00){\line(3,-5){15.00}}
\put(95.00,-10.00){\line(3,5){15.00}}
\put(95.00,-10.00){\line(1,0){30.00}}
\put(117.00,-4.00){\vector(-2,3){5.33}}
\put(-5.00,134.00){\makebox(0,0)[cc]{(a)}}
\put(-5.00,67.00){\makebox(0,0)[cc]{(b)}}
\put(-5.00,4.00){\makebox(0,0)[cc]{(c)}}
\put(31.00,90.00){\vector(2,-3){5.30}}
\put(10.00,105.00){\line(1,0){10.00}}
\put(90.00,105.00){\line(-1,0){10.00}}
\put(80.00,105.00){\line(0,0){0.00}}
\put(38.00,39.00){\circle{6.00}}
\put(30.00,52.00){\circle{6.00}}
\put(22.00,39.00){\circle{6.00}}
\put(30.00,52.00){\makebox(0,0)[cc]{F}}
\put(22.00,39.00){\makebox(0,0)[cc]{F}}
\put(38.00,39.00){\makebox(0,0)[cc]{E}}
\put(30.00,60.00){\line(3,-5){15.00}}
\put(45.00,35.00){\line(-1,0){30.00}}
\put(15.00,35.00){\line(3,5){15.00}}
\put(29.00,50.00){\vector(-2,-3){5.33}}
\put(78.00,124.00){\circle{6.00}}
\put(70.00,137.00){\circle{6.00}}
\put(62.00,124.00){\circle{6.00}}
\put(70.00,137.00){\makebox(0,0)[cc]{F}}
\put(62.00,124.00){\makebox(0,0)[cc]{F}}
\put(78.00,124.00){\makebox(0,0)[cc]{X}}
\put(70.00,145.00){\line(3,-5){15.00}}
\put(85.00,120.00){\line(-1,0){30.00}}
\put(55.00,120.00){\line(3,5){15.00}}
\put(69.00,135.00){\vector(-2,-3){5.33}}
\put(77.00,39.00){\circle{6.00}}
\put(69.00,52.00){\circle{6.00}}
\put(61.00,39.00){\circle{6.00}}
\put(69.00,52.00){\makebox(0,0)[cc]{F}}
\put(61.00,39.00){\makebox(0,0)[cc]{F}}
\put(77.00,39.00){\makebox(0,0)[cc]{E}}
\put(69.00,60.00){\line(3,-5){15.00}}
\put(84.00,35.00){\line(-1,0){30.00}}
\put(54.00,35.00){\line(3,5){15.00}}
\put(68.00,50.00){\vector(-2,-3){5.33}}
\put(24.00,39.00){\vector(1,0){10.00}}
\put(31.00,50.00){\vector(2,-3){5.33}}
\put(90.00,105.00){\line(0,-1){10.00}}
\put(90.00,83.00){\line(0,-1){10.00}}
\put(90.00,30.00){\line(0,1){10.00}}
\put(90.00,52.00){\line(0,1){10.00}}
\put(68.00,105.00){\line(-1,0){10.00}}
\put(42.00,105.00){\line(-1,0){10.00}}
\put(10.00,30.00){\line(1,0){10.00}}
\put(90.00,30.00){\line(-1,0){10.00}}
\put(68.00,30.00){\line(-1,0){10.00}}
\put(42.00,30.00){\line(-1,0){10.00}}
\put(10.00,105.00){\line(0,-1){10.00}}
\put(10.00,83.00){\line(0,-1){10.00}}
\put(10.00,30.00){\line(0,1){10.00}}
\put(10.00,52.00){\line(0,1){10.00}}
\put(72.00,-30.0){\makebox(0,0)[cc]{Fig. 2.}}
\end{picture}
\end{figure}

\newpage
\begin{thebibliography}{99}
\bibitem{BTW} P.~Bak, C.~Tang and K.~Wiesenfeld, Phys. Rev. Lett. 59 (1987)
381; Phys. Rev. A 38 (1988) 364.
\bibitem{PVZ} L.~Pietronero, A.~Vespignani and S.~Zapperi, Phys. Rev. Lett.
72 (1994) 1690; A.~Vespignani, S.~Zapperi and L.~Pietronero, Phys. Rev. E 51
(1995) 1711.
\bibitem{LPVZ} V.~Loreto, L.~Pietronero, A.~Vespignani and S.~Zapperi,
Phys. Rev. Lett. 75 (1995) 465.
\bibitem{EVI} E.V.~Ivashkevich, Phys. Rev. Lett. 76 (1996) 3368.
\bibitem{DD} D.~Dhar, Phys. Rev. Lett. 64 (1990) 1613.
\bibitem{VBP} V.B.~Priezzhev, J. Stat. Phys. 74 (1994) 955.
\bibitem{PKI} V.B.~Priezzhev, D.V.~Ktitarev and E.V.~Ivashkevich, Phys.
Rev. Lett. 76 (1996) 2093.
\bibitem{MD1} S.N.~Majumdar, D.~Dhar, J. Phys. A 24 (1991) 475L.
\bibitem{MD2} S.N.~Majumdar, D.~Dhar, Physica A 185 (1992) 129.
\bibitem{MDM} S.S.~Manna, D.~Dhar and S.N.~Majumdar, Phys. Rev. A 46 (1992)
R4471.
\end{thebibliography}
\end{document}